# Reliability Analysis of Component Level Redundant Topologies for solid state Fault Current Limiter

Masoud Farhadi [1], Mehdi Abapour [2*], Behnam Mohammadi-Ivatloo [3]

[1] Department of Electrical and Computer Engineering, University of Tabriz, 29 Bahman Blvd. Tabriz, Iran
[2] Department of Electrical and Computer Engineering, University of Tabriz, 29 Bahman Blvd. Tabriz, Iran
[3] Department of Electrical and Computer Engineering, University of Tabriz, 29 Bahman Blvd. Tabriz, Iran
[*] corresponding.abapour@tabrizu.ac.ir

**Abstract:** Experience shows that semiconductor switches in power electronics systems are the most vulnerable components. One of the most common ways to solve this reliability challenge is component-level redundant design. There are four possible configurations for the redundant design in component-level. This paper presents a comparative reliability analysis between different component-level redundant designs for solid state fault current limiter (SSFCL). The aim of the proposed analysis is to determine the more reliable component-level redundant configuration. The mean time to failure (MTTF) is used as the reliability parameter. Considering both fault types (open circuit and short circuit), the MTTFs of different configurations are calculated. It is demonstrated that more reliable configuration depends on the junction temperature of the semiconductor switches in the steady state. That junction temperature is a function of i) ambient temperature, ii) power loss of the semiconductor switch and iii) thermal resistance of heat sink. Also, results' sensitivity to each parameter is investigated. The results show that in different conditions, various configurations have higher reliability. The experimental results are presented to clarify the theory and feasibility of the proposed approaches. At last, levelized costs of different configurations are analyzed for a fair comparison.

## Nomenclature

| | |
|---|---|
| $\lambda_{O.C}$ | Open circuit rate of unidirectional switch (failures per million calendar hours or FIT, 1 FIT=$10^{-9}$ failure/hour). |
| $\lambda_{S.C.}$ | Short circuit rate of unidirectional switch (FIT). |
| $\lambda_{O.C,H}$ | Open circuit rate of unidirectional switch at half load (FIT). |
| $\lambda_{S.C,H}$ | Short circuit rate of unidirectional switch at half load (FIT). |
| $P_S$ | Perfect relaying probability. |
| $P_{SH,P}$ | Perfect relaying probability in shunt redundant configuration with parallel operation from a reliability point of view. |
| $P_{S,SB}$ | Perfect relaying probability in series redundant configuration with standby operation from a reliability point of view. |
| $\lambda_{SW}$ | Failure rate of bidirectional switch (FIT). |
| $\lambda_{SW,H}$ | Failure rate of bidirectional switch at half load (FIT). |
| $\lambda_S$ | Failure rate of unidirectional switch (FIT). |
| $MTTF_{SH,P}$ | MTTF of SSFCL with shunt redundant switch (parallel from a reliability point of view) ($10^6$ h). |
| $MTTF_{SH,SB}$ | MTTF of SSFCL with shunt redundant switch (standby from a reliability point of view) ($10^6$ h). |
| $MTTF_{S,P}$ | MTTF of SSFCL with series redundant switch (parallel from a reliability point of view) ($10^6$ h). |
| $MTTF_{S,SB}$ | MTTF of SSFCL with series redundant switch (standby from a reliability point of view) ($10^6$ h). |
| $\lambda_b$ | Base failure rate (FIT). |
| $\pi_T$ | Temperature factor. |
| $\pi_{T,H}$ | Temperature factor at half load. |
| $\pi_Q$ | Quality factor. |
| $\pi_E$ | Environment factor. |
| $T_j$ | Junction temperature (°C). |
| $T_{j,H}$ | Junction temperature at half load (°C). |
| $T_a$ | Ambient temperature (°C). |
| $T_C$ | Case temperature (°C). |
| $T_H$ | Heat sink temperature (°C). |
| $R_{th,ca}$ | Heat sink thermal resistance (°C/W). |
| $R_{jC}$ | Thermal resistance between the junction and case (°C/W). |
| $R_{th,cH}$ | Thermal resistance between the case and heat sink (°C/W). |
| $R_{th,Ha}$ | Thermal resistance between the heat sink and ambient (°C/W). |
| $P_{loss}$ | Power losses (W). |
| $P_{loss,H}$ | Power losses at half load (W). |
| $P_{SW}$ | Switching losses (W). |
| $f_{SW}$ | Switching frequency (Hz). |
| $P_{cond}$ | Conduction losses (W). |
| $E_{cond}$ | Conduction energy (J). |
| $E_{on}$ | Turn on energy loss (J). |
| $E_{off}$ | Turn off energy loss (J). |
| $V_0$ | Threshold voltage (V). |
| $i(t)$ | Instantaneous switch current (A). |
| $R_S$ | Static resistance of switch (Ω). |
| $T_S$ | Sampling period (Sec). |

## 1. Introduction

Nowadays, increasing consumption demand and continuous growth of distributed generation (DG), has led to short circuit level increment in interconnected networks. This exceeded current is one of the main concerns of network operators. In recent years, various solutions are proposed to solve this problem. Considering their technical, political and economic difficulties, the use of fault current limiter (FCL) is one of the most promising proposed methods [1]-[2]. Up to now, different studies have been carried out on FCLs with specifically focused on their impacts on reliability of faulty systems [1]–[3]. But, issues like FCL reliability calculations or fault-tolerant design are not discussed, while reliability of FCL has a large effect on its impact on system reliability.

Recently, various solutions have been proposed for fault-tolerant design. These solutions could be classified into two major categories: 1) algorithmic solutions without change of the existing hardware [4]-[5]; and 2) fault-tolerant designs with redundant hardware [6]-[8]. The second category could be also divided into four categories: a) component-level, b) leg-level, c) module-level, and d) system-level [9]. Component-level redundancy, in general, is the duplication of extra switches in order to have a backup in the case of post fault [10], [11]. Leg-level redundancy is provided by incorporating extra legs in parallel or series connection into the main legs. Extra leg is added to override the effects of a failed leg. Module-level redundancy can be further subdivided into: i) neutral shift, ii) DC-bus voltage reconfiguration and iii) redundant module installation. Neutral-shift strategy attempts to adjust phase shifts among phase-voltage references to maintain balanced line-to-line voltages in post fault [12]-[13]. The second strategy involves attempts to sustain an unchanged output voltage by raising the input voltage [14]. Redundant module installation, in general, is the duplication of extra modules in order to have a backup in the case of post fault [15]. System-level redundancy is the most expensive redundancy that can be employed in industrial applications. Two usual types of system-level redundancy are the cascaded redundant and the parallel redundant [16]-[17]. Among hardware redundancy solutions, the component-level redundant solution achieves a better compromise between the system cost and simplicity that has become a hot area of fault-tolerant research.

In [18], the authors present a global reliability comparison between two-level and three-level/five-level inverter topologies in single and three-phase operations. Harb et al. has proposed a new methodology for calculating the reliability of the photovoltaic module-integrated inverter (PV-MII) based on a stress factor approach [19]. In [20], the reliability analysis of the power electronic converters for a grid connected permanent magnet generator-based wind energy conversion system is presented. In [21], in this paper FIDES method has been governed to estimate the reliability of two industrial AC/DC converters with resonant and non-resonant topologies. In [22], a study of different inverter topologies to increase reliability and avoid an expensive redundancy is presented. Many points have been considered and discussed for performance of parallel and series connected switches in previous researches. However, the reliability point of view has not been studied.

For the first time to our knowledge, in this paper, a comparative reliability analysis between different component-level redundant topologies for FCL has been proposed. Though the methodology presented is general and can be extended to other similar power electronics systems, results associated with a solid state fault current limiter (SSFCL) are presented and discussed in this work. Two different operation scenarios (perfect fault coverage and imperfect fault coverage) are considered and more reliable configurations are determined for different operation scenarios.

In order to ensure increasing current and voltage rating of power switches, paralleling and serializing is inevitable. On the other hand, it is a technical challenging task to ensure proper current and voltage sharing between the parallel and series connected semiconductor switches. Control strategies have been suggested in [23]–[25] to improve the dynamic and static sharing between switches. It should be noted that prior to this, the voltage and current ratings of commercially available semiconductors were limited and far below high voltage application requirements. Therefore, the suitable topology was selected based on available ratings of semiconductors. But today, currently available semiconductors is able to endure surge current up to 80 kA. In consequence, reliability challenges should be considered in power electronics circuit design.

It is demonstrated that the more reliable configuration is determined by the junction temperature of the switches. This is because of high contribution of temperature factor in failure rate of power semiconductor devices. The junction temperature is a function of ambient temperature, power loss of the semiconductor switch, and thermal resistance of heat sink. By controlling these parameters, we can achieve an appropriate configuration with more reliability.

## 2. Length state space diagrams and MTTFs of different redundant topologies for SSFCL

SSFCL topologies are classified into three major groups: the series switch, the bridge, and the resonant types [26]. In this paper, without loss of generality, a generic topology of the series switch-type FCL is selected for discussion and analysis. The topology of series switch-type FCL incorporated in a single phase power line is shown in Fig. 1. It consists of a bidirectional AC switch, a fault current limiting inductor $L_M$, and a voltage limiting element (e.g. MOV: metal oxide varistor). The principle of series switch-type FCL operation has been verified by experiments and simulations [19]. The field experience confirms that power switches are the most vulnerable components. Moreover, magnetic components and MOVs are much more reliable [20]-[22]. Therefore, only power switches are considered in component level redundancy. Note that it is assumed the SSFCL is non-repairable, capable to short/open circuit detection, isolation and reconfiguration (DIR). Also, it is assumed that the bidirectional switch is a pair of anti-parallel switches. To observe abridgement in subsequent text, switch will be used instead of bidirectional switch. In the following, the state space diagrams and MTTFs for four possible redundant configurations in component level are explained that can provide a useful starting point.

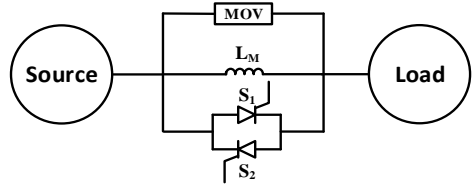

*Fig. 1.* *Generic topology of the series switch-type FCL incorporated in a single-phase power line.*

*2.1. SSFCL with shunt redundant switch (Parallel from a reliability point of view)*

In this configuration, main and redundant switches are paralleled and each switch is in series with a low frequency electromechanical relay. Both relays are normally-closed and in case of fault detection in each switch, respective relay will be open. In this case, the requirement for system failure is that both switches fail, so switches are connected in parallel from a reliability point of view. This configuration is shown in Fig. 2a. In this case, each switch carries half of the load current. After fault occurrence, the DIR controller opens the corresponding relay and total current will be carried by intact switch and this leads to differences in failure rates before and after the fault. Therefore we will use the subscript "H" for failure rate before the fault. Fig. 2b shows the state space diagram of this configuration. The system has seven performance states and each status is shown by a rectangle. The failure rates corresponding to a transition from state i to state j are shown on the corresponding arrows. For example, failure rate corresponding to a transition from state 1 to state 3 is equal to multiplication of short-circuit rate of, one of two switches in half-load $2\lambda_{S.C,H}$, and perfect relaying probability $P_S$. If the relay cannot open, FCL is short circuited. So the failure rate corresponding to a transition from state 1 to state 4 is equal to multiplication of short-circuit rate of one of two switches in half-load $2\lambda_{S.C,H}$, and relay failure probability $(1-P_S)$. Now, using state space diagram we can form stochastic transitional probability matrix, P. P is an n-by-n square matrix for an n-state system. Where $P_{ij}$ (i ≠ j) is transition probability from state i to state j. Since the summation of the probabilities in each row of the P is equal to one, so $P_{ii}$ is $(1-\sum P_{ij})$. The stochastic transitional probability matrix for this case is as follows:

$$P = \begin{bmatrix} 1-2(\lambda_{O.C,H}+\lambda_{S.C,H}) & 2\lambda_{O.C,H} & 2P_S\lambda_{S.C,H} & 2(1-P_S)\lambda_{S.C,H} & 0 & 0 & 0 \\ 0 & 1-(\lambda_{OC}+\lambda_{SC}) & 0 & 0 & \lambda_{OC} & \lambda_{SC} & 0 \\ 0 & 0 & 1-(\lambda_{OC}+\lambda_{SC}) & 0 & 0 & \lambda_{OC} & \lambda_{SC} \\ 0 & 0 & 0 & 1 & 0 & 0 & 0 \\ 0 & 0 & 0 & 0 & 1 & 0 & 0 \\ 0 & 0 & 0 & 0 & 0 & 1 & 0 \\ 0 & 0 & 0 & 0 & 0 & 0 & 1 \end{bmatrix} \quad (1)$$

Then, by defining the absorbing states as a state which leads to system failure, truncated probability matrix, $Q$, is formed based on $P$ and by deleting the rows and columns associated with the absorbing states (For further information, refer [23]–[25]). According to the state space diagram in this configuration, states 4-7 are absorbing states which, once entered, cannot be left until the process starts again. So the truncated matrix $Q$ is given by:

$$Q = \begin{bmatrix} 1-2(\lambda_{O.C,H}+\lambda_{S.C,H}) & 2\lambda_{O.C,H} & 2P_S\lambda_{S.C,H} \\ 0 & 1-(\lambda_{OC}+\lambda_{SC}) & 0 \\ 0 & 0 & 1-(\lambda_{OC}+\lambda_{SC}) \end{bmatrix} \quad (2)$$

According to the (3) and (4), truncated matrix can be used to calculate the average time which is expected to last in operation, (i.e., MTTF), where $M$ and $I$ are the fundamental matrix and unit matrix, respectively.

$$M = [I - Q]^{-1}_{3\times 3} \quad (3)$$

If the system starts in state i, the MTTF of the system is sum of the values in ith row of fundamental matrix, $M$. So, with assumption that the system starts in state 1, the MTTF is:

$$MTTF_{SH,P} = \sum_{i=1}^{3} m_{1,i} = \frac{\lambda_{O.C}+\lambda_{S.C}+2(\lambda_{O.C,H}+P_S\lambda_{S.C,H})}{2(\lambda_{O.C}+\lambda_{S.C})(\lambda_{O.C,H}+\lambda_{S.C,H})} \quad (4)$$

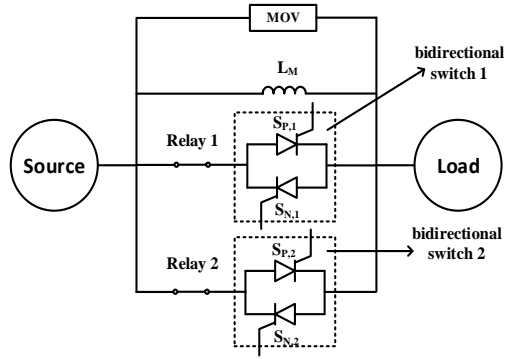

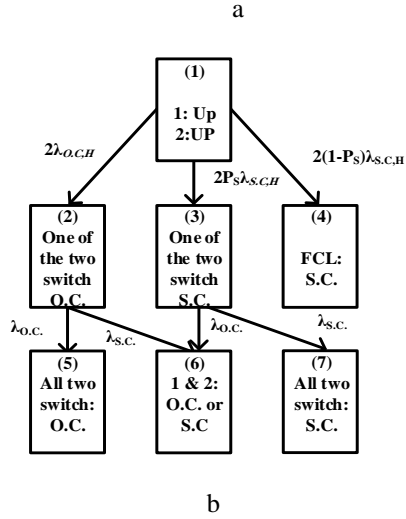

a

b

**Fig. 2.** *Configuration and state space diagram of the SSFCL with shunt redundant switch (Parallel from a reliability point of view)*

a Configuration

b State space diagram

### 2.2. SSFCL with shunt redundant switch (Standby from a reliability point of view)

In this configuration switches and relays are connected similar to the previous case, but corresponding relay to the main switch and corresponding relay to the redundant switch will be normally-closed and normally-open, respectively. In this case redundant switch will enter to circuit after fault occurrence in main switch, so switches are connected in standby from a reliability point of view that leads each of switches to carry full load current. After fault detection, corresponding relay to the main switch is open and corresponding relay to the redundant switch is closed to isolate fault and reconfiguration. At first sight, it may be thought that this configuration is more reliable than the previous configuration, but in following we will see that this statement is not always true.

Fig. 3 show the configuration of the SSFCL with shunt redundant switch (standby from a reliability point of view) and corresponding state space diagram. Using state space diagram and procedure outlined in the previous section, MTTF in this case is calculated as follow:

$$MTTF_{SH,SB} = \frac{\lambda_{O.C} + \lambda_{S.C} + P_S \lambda_{O.C} + P_S^2 \lambda_{S.C}}{(\lambda_{O.C} + \lambda_{S.C})^2} \quad (5)$$

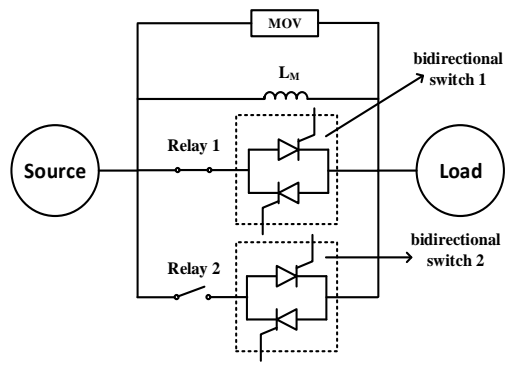

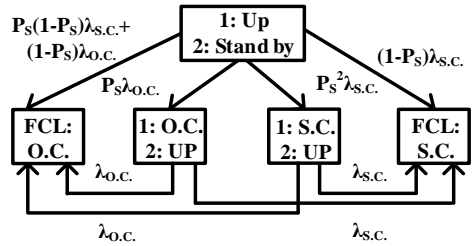

*Fig. 3. Configuration and state space diagram of the SSFCL with shunt redundant switch (Standby from a reliability point of view)*

a Configuration

b State space diagram

### 2.3. SSFCL with series redundant switch (Parallel from a reliability point of view)

In this configuration main and redundant switches are in series together and each switch is paralleled with a relay. Both relays are normally-open and after fault detection in each switch, respective relay will be short circuit. This configuration is parallel from a reliability point of view. As noted above, the bidirectional switch is a pair of anti-parallel switches. In each of these switches, the applied voltage is much less than rated voltage (applied voltage /rated voltage < 0.3), so according to [26], the voltage stress factor will always be the lowest value. On the other hand, regardless of how many switches are in the circuit, each of switches carries full load current; therefore in this configuration there is no difference between failure rates before and after the first fault. Also from the beginning, both of switches are in the circuit.

Fig. 4 shows the configuration of the SSFCL with series redundant switch (parallel from a reliability point of view) and corresponding state space diagram. The MTTF of this configuration is calculated as follows:

$$MTTF_{S,P} = \frac{\lambda_{O.C} + \lambda_{S.C} + 2\lambda_{S.C} + 2P_S \lambda_{O.C}}{2(\lambda_{O.C} + \lambda_{S.C})^2} \quad (6)$$

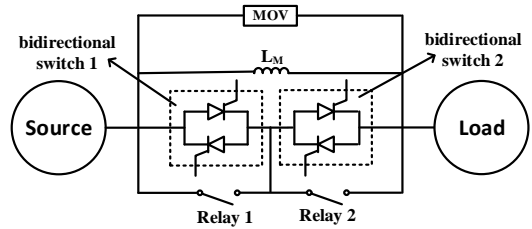

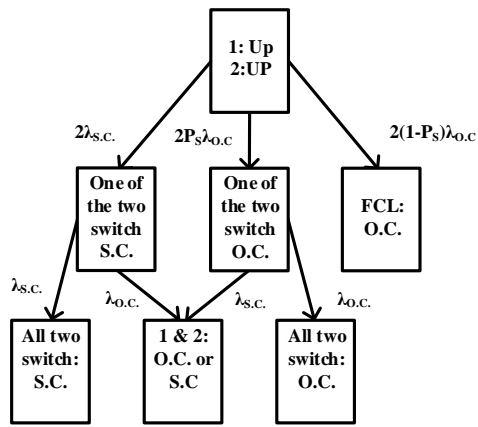

*Fig. 4. Configuration and state space diagram of the SSFCL with series redundant switch (Parallel from a reliability point of view)*

a Configuration

b State space diagram

### 2.4. SSFCL with series redundant switch (Standby from a reliability point of view)

In this configuration main and redundant switches are in series; corresponding relay to the main switch and corresponding relay to the redundant switch will be normally-open and normally-closed, respectively. In this case, redundant switch will enter to circuit after fault occurrence in main switch, so switches are connected in standby from a reliability point of view. This configuration is like second configuration with difference that, this time redundant switch enters to circuit that is in series with main switch circuit. So, if the probability of fault coverage, $P_S$ is one; expected MTTFs of these configurations are same as together. The main role of relays in this configuration is open circuit fault isolation. This means that if a short circuit occurs in each switch; its corresponding relay does not operate, while the main role of relays in second configuration is short circuit fault isolation. Since short circuit faults impose the most power switches faults, so if the probability of fault coverage, $P_S$ is less than one; and this configuration is more reliable than the second configuration.

On the other hand, if a short circuit fault occurs in main switch; to isolate this fault in the second configuration, corresponding relay to the main switch and corresponding relay to the redundant switch must be open and closed, respectively. But in this configuration only corresponding relay to the redundant switch is opened. The failure rate of the relay contacts is composed by two failure modes: the failure in closing the contacts and the failure in opening the contacts. The contribution of the failure in closing the contacts is often assumed to be some fifty times greater than the contribution of the failure in opening the contacts [27]. Again this configuration is more reliable than the second configuration.

Fig. 5 show the configuration of the SSFCL with series redundant switch (standby from a reliability point of view) and corresponding state space diagram. The MTTF of this configuration is calculated as follows:

$$MTTF_{S,SB} = \frac{\lambda_{o.c} + \lambda_{s.c} + P_S^2 \lambda_{o.c} + P_S \lambda_{s.c}}{(P_S^2 \lambda_{o.c} + \lambda_{s.c})(\lambda_{o.c} + \lambda_{s.c})} \quad (7)$$

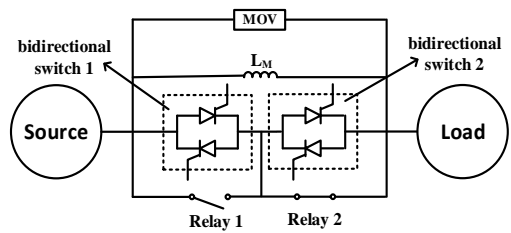

a

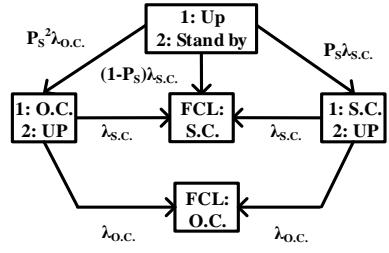

b

*Fig. 5.* Configuration and state space diagram of the SSFCL with series redundant switch (Standby from a reliability point of view)

a Configuration

b State space diagram

## 3. Reliability Comparison of Different SSFCL Redundant Topologies in Perfect Fault Coverage

Since, relays are mission oriented components in FCL (not continuously operated components), the information that is given in the empirical-based data sources is not suitable for relays. So in this paper, fault coverage (fault detection, isolation, and reconfiguration) probability ($P_S$) has been employed. To compare different topologies in this paper, both different operation scenarios (perfect fault coverage and imperfect fault coverage) are considered. In this section, a comparative reliability study is carried out to determine the more reliable configuration in perfect fault coverage. So, the probability of fault coverage is assumed unity. Also, sum of open circuit rate and short circuit rate, is defined as failure rate (i.e. $\lambda_{O.C} + \lambda_{S.C.} = \lambda_{SW}$), which is equal to twice as the failure rate of a unidirectional switch ($\lambda_{SW} = 2\lambda_S$). Considering these assumptions, the MTTFs equations can be rewritten as follows:

$$MTTF_{SH,P} = \frac{\lambda_{SW} + 2\lambda_{SW,H}}{2\lambda_{SW}\lambda_{SW,H}} \quad (8)$$

$$MTTF_{SH,SB} = MTTF_{S,SB} = \frac{2}{\lambda_{SW}} \quad (9)$$

$$MTTF_{S,P} = \frac{3}{2\lambda_{SW}} \quad (10)$$

Comparing (8)-(10), it is clear that the series redundant configuration with parallel operation from a reliability point of view has the lowest MTTF. Also it can be seen that standby topologies from a reliability point of view have the same MTTF. Now we compare the shunt redundant configuration (parallel from a reliability point of view) with standby configurations from a reliability point of view. To do this the following inequality is used. MTTFs of these configurations as a function of switch failure rate in full load and half load are compared in Fig. 6.

$$MTTF_{SH,P} > MTTF_{SH,SB} = MTTF_{S,SB} \quad (11)$$

$$\frac{\lambda_{SW} + 2\lambda_{SW,H}}{2\lambda_{SW}\lambda_{SW,H}} > \frac{2}{\lambda_{SW}} \quad (12)$$

This implies that

$$\frac{\lambda_{SW}}{\lambda_{SW,H}} > 2 \quad (13)$$

According to the MIL-HDBK-217 [26], as major reference of failure rate calculation, the general form of failure rate of power semiconductor switch (unidirectional switch) is calculated as (14), that other parameters of (14) is same for all configurations except $\pi_T$. The general form of $\pi_T$, is as (15):

$$\lambda_S = \lambda_b \pi_T \pi_Q \pi_E \prod_i \pi_i \quad (14)$$

$$\pi_T = e^{-a\left(\frac{1}{T_j+273}-\frac{1}{298}\right)} \quad (15)$$

From (13)–(15), following equations can be obtained:

$$\frac{\pi_T}{\pi_{T,H}} > 2 \quad (16)$$

$$\frac{1}{T_{j,H}+273} - \frac{1}{T_j+273} > \frac{\ln(2)}{a} \quad (17)$$

$$T_j > \frac{\left(1+\frac{\ln(2)}{a}273\right)T_{j,H} + \frac{\ln(2)}{a}273^2}{-\frac{\ln(2)}{a}T_{j,H} + \left(1-\frac{\ln(2)}{a}273\right)} \quad (18)$$

Regardless of the term $\{\ln(2)T_{j,H}/a\}$, the condition in which the shunt redundant configuration with parallel operation from a reliability point of view is more reliable can be derived as:

$$T_j > \frac{(a+273\ln(2))T_{j,H}+\ln(2)273^2}{a-273\ln(2)} = C_1 T_{j,H} + C_2 \quad (19)$$

With the assumption that the bidirectional switch is composed of a pair of thyristors, a, $C_1$, $C_2$ coefficients in (19) are 3082, 1.1308, and 17.8582, respectively. Boundary condition, in which both shunt redundant configuration (parallel from a reliability point of view) and redundant configuration with standby operation from a reliability point of view have the same MTTF, is shown in Fig. 7, which regions (a) and (b) are region in which the shunt redundant switch (parallel from a reliability point of view) configuration and region in which the redundant switch (standby from a reliability point of view) configuration are more reliable, respectively.

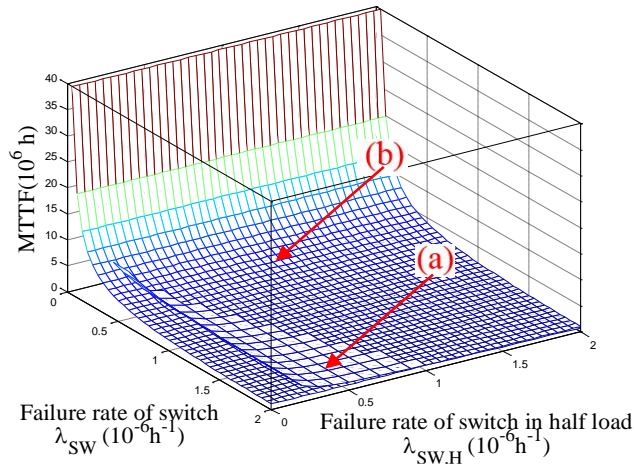

***Fig.6.*** *MTTF of the SSFCL as a function of failure rate of switch at full load and half load. (a) Shunt redundant configuration (parallel from a reliability point of view). (b) Series redundant configuration (standby from a reliability point of view).*

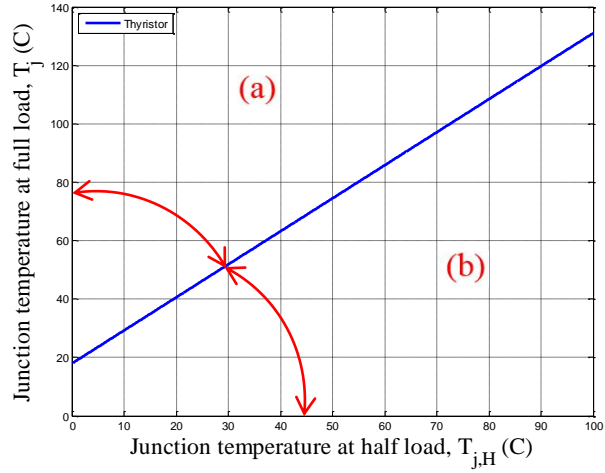

***Fig.7.*** *MTTF of the SSFCL as a function of failure rate of switch at full load and half load and Boundary condition. (a) Region in which the shunt redundant configuration (parallel from a reliability point of view) is more reliable. (b) Region in which the redundant configuration with standby operation from a reliability point of view is more reliable.*

## 4. Reliability Comparison Between Different SSFCL Redundant Topologies in Imperfect Fault Coverage

In this scenario, it is assumed that the fault coverage (fault detection, isolation, and reconfiguration) is imperfect ($P_S \neq 1$). As mentioned in Section 2, in this case, the series redundant configuration (standby from a reliability point of view) has a higher MTTF than the shunt redundant configuration (standby from a reliability point of view). So in this section, series redundant configuration (standby from a reliability point of view) and shunt redundant configuration (parallel from a reliability point of view) are compared. Probabilities in different cases must be segregated. Also, the percentages of the different failure modes should be specified. If we assume that $\chi\%$ of the faults are short circuit; we can write the following equations:

$$\lambda_{S.C} = \chi . \lambda_{SW} \quad (20)$$

$$\lambda_{O.C} = (1-\chi) . \lambda_{SW} \quad (21)$$

Considering this assumption, the MTTFs equations can be rewritten as follows:

$$MTTF_{SH,P} = \frac{\lambda_{SW} + [2 + 2\chi(P_{SH,P} - 1)]\lambda_{SW,H}}{2\lambda_{SW}\lambda_{SW,H}} \quad (22)$$

$$MTTF_{S,SB} = \frac{1 + P_{S,SB}^2 + (P_{S,SB} - P_{S,SB}^2)\chi}{\lambda_{SW}[P_{S,SB}^2 + (1 - P_{S,SB}^2)\chi]} \quad (23)$$

Using these equations, the condition in which the shunt redundant configuration with parallel operation from a reliability point of view is more reliable can be derived as:

$$\frac{\lambda_{SW,H}}{\lambda_{SW}} < \frac{0.5[\chi + (1-\chi)P_{S,SB}^2]}{1 - \chi + \chi^2 + \chi P_{S,SB} - \chi^2 P_{SH,P} + (\chi - \chi^2)P_{S,SB}^2(1 - P_{SH,P})} \quad (24)$$

The last term in the denominator of (24), i.e., $(\chi-\chi^2).P_{S,SB}^2.(1-P_{SH,P})$ can be safely neglected.

$$\frac{\lambda_{SW,H}}{\lambda_{SW}} < \frac{0.5[\chi + (1-\chi)P_{S,SB}^2]}{(1 - \chi + \chi^2) + \chi P_{S,SB} - \chi^2 P_{SH,P}} \quad (25)$$

Where, $\gamma$ is defined as the ratio of the perfect relaying probability in shunt redundant configuration to perfect relaying probability in series redundant configuration which can have a value between zero and one and is usually close to one. Also, according to [28], the percentage of the short circuit failure, $\chi$ is 0.98 for thyristor. Thus, following equations can be deduced:

$$P_{S,SB} = \gamma P_{SH,P} \quad (26)$$

$$\left(\frac{\lambda_{SW,H}}{\lambda_{SW}}\right)_{Thyristor} < \frac{0.5(0.98 + 0.02\gamma^2 P_{SH,P}^2)}{0.98 + (0.98\gamma - 0.96)P_{SH,P}} \quad (27)$$

On other hand, according to the MIL-HDBK-217, failure rate ratio based on temperature factor ratio is calculated as follows:

$$\frac{\lambda_{SW,H}}{\lambda_{SW}} = \frac{\pi_{T,H}}{\pi_T} = e^{a(\frac{1}{T_j + 273} - \frac{1}{T_H + 273})} \quad (28)$$

Now, with comparison of (27) and (28), we can identify the appropriate redundant configuration. In Fig. 8, (27) and (28) have been cut and appropriate regions for shunt redundant configuration (parallel from a reliability point of view) and series redundant configuration (standby from a reliability point of view) are determined.

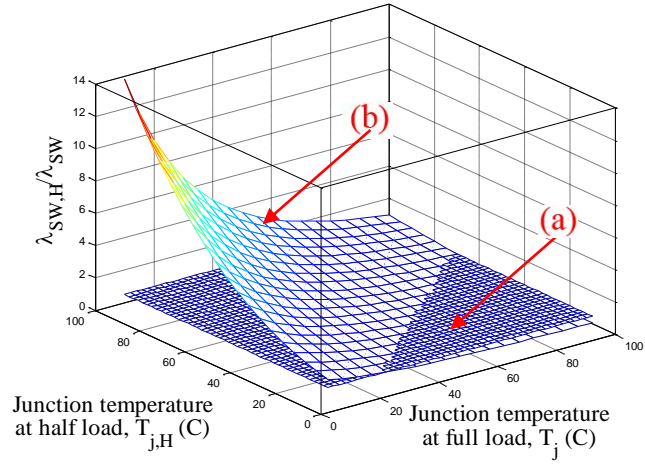

*Fig. 8. Failure rate of SSFCL at half load to full load ratio. (a) Region in which the shunt redundant configuration (parallel from a reliability point of view) is more reliable. (b) Region in which the series redundant configuration (standby from a reliability point of view) is more reliable.*

## 5. Sensitivities of More Reliable Configuration to Different Parameters

In this section, sensitivities of more reliable configuration to ambient temperature, power losses, and thermal resistance of heat sink are analyzed. As seen in the previous section, predicted lifetime of each configuration is a function of junction temperature of switches. Also, more reliable configuration has been determined by junction temperature range, therefore junction temperature is one of the principal parameters and a common input parameter of the reliability calculation. The used thermal model for power switch is shown in Fig. 9a, which the thermal impedance between the junction and case, is usually modeled as a multi-layers foster RC network in the manufacturer datasheets (see Fig. 9b). Regardless of the thermal capacitance $C_{th}$, which describes dynamic changes, the junction temperature based on the thermal equivalent model of Fig.9a, is calculated as follows:

$$T_j = T_a + P_{loss}(R_{th,jC} + R_{th,Ca}) \quad (29)$$

Usually sum of thermal resistance between the case-heat sink and thermal resistance between the heat sink-ambient is considered as the thermal resistance of heat sink.

$$R_{th,Ca} = R_{th,CH} + R_{th,Ha} \quad (30)$$

From (19), (29), and (30) region which the shunt redundant configuration with parallel operation from a reliability point of view is more reliable, can be derived as:

$$P_{loss} - C_1 P_{loss,H} > \frac{(C_1-1)T_a + C_2}{R_{th,jC} + R_{th,Ca}} \quad (31)$$

Since the ambient temperature coefficient is positive in (31), so it is clear that decreasing the ambient temperature leads (31) more likely to be satisfied. Also, with increasing the rated power level, ($P_{loss}$-$C_1P_{loss,H}$) -the left term- increases as well. Therefore increasing the rated power level leads the above inequality more likely to be satisfied. In other words, the shunt redundant configuration with parallel operation from a reliability point of view is more appropriate for the systems with high power and low ambient temperature applications. However, for an application with specified ambient temperature and rated power level, thermal resistance of heat sink is the main parameter to determine the more reliable configuration.

Fig. 9c compares MTTFs of two candidate topologies (i.e. shunt redundant configuration with parallel operation from a reliability point of view and series redundant configuration with standby operation from a reliability point of view) as a function of failure rate of switch in full load and half load.

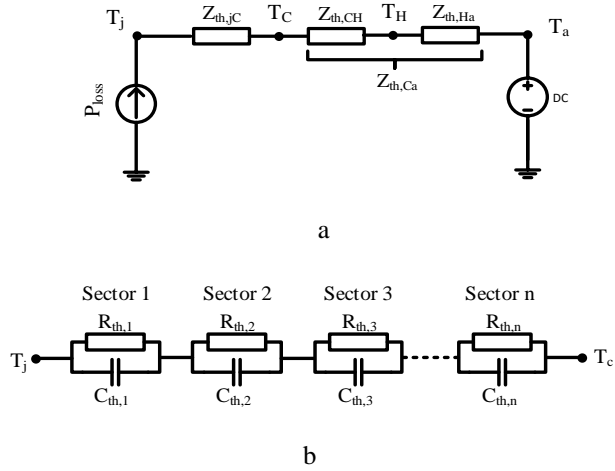

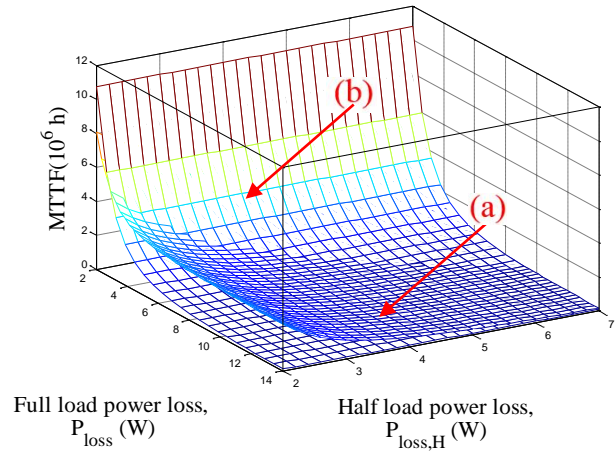

c

*Fig. 9.* *The used thermal model, thermal equivalent and MTTF of the SSFCL*

a The used thermal model for thyristor module.

b Thermal equivalent for $Z_{th,jC}$ (Foster network, usually n=4).

c MTTF of the SSFCL as a function of full load power losses and half load power losses. (a) Shunt redundant configuration (parallel from a reliability point of view). (b) Series redundant configuration (standby from a reliability point of view).

## 6. Experimental Results

To have a better understanding of (19), and (31), a case study has been performed in this section. To do this, a low-power solid state fault current limiter (SSFCL) with shunt redundant configuration (parallel from a reliability point of view) has been developed. Since in the standby operation, only one switch is in the circuit, and purpose of this section is to calculate power losses and junction temperature of switches. Thus instead of series redundant configuration, we use a SSFCL without redundant component. Fig. 10a shows the experimental setup of the SSFCL with shunt redundant configuration (parallel from a reliability point of view) which is able to lead out redundant switch. The circuit and SSFCL parameters can be found in Table 2 in the Appendix.

Fig. 10 shows thermal image of two mentioned topologies in steady state conditions which instead of the inaccessible junction temperatures, represent case temperatures of the individual switches. The case temperature of thyristors in each topology is given in the images (i.e. $44.46\,^{\circ}\text{C}$, and $104.46\,^{\circ}\text{C}$ for shunt redundant configuration and without redundant configuration, respectively). According to Fig. 10 and (29), the junction temperature of switches in shunt redundant configuration and without redundant configuration are 44.90$^{\circ C}$, and 106.20$^{\circ C}$, respectively. Also according to the procedures specified in Appendix, the power loss of thyristors is calculated. The power losses of each thyristor in shunt redundant configuration and without redundant configuration are 0.335$^W$, and 1.34$^W$, respectively. Based on obtained values for junction temperature and power losses (19), and (31) are satisfied. So, shunt redundant configuration (parallel from a reliability point of view) is more reliable for SSFCL with the given properties.

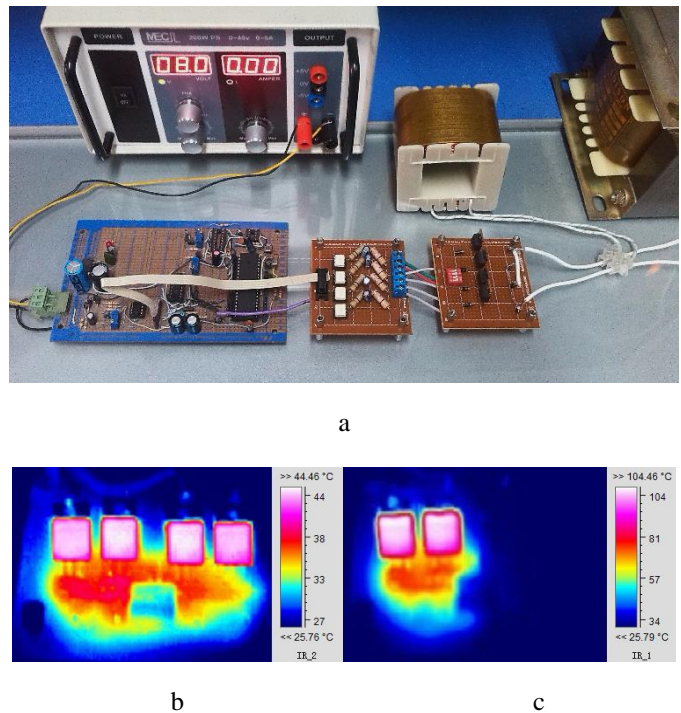

*Fig. 10. Laboratory prototype and thermal images of the SSFCL.*

a Laboratory prototype

b Thermal images of the SSFCL with shunt redundant configuration (parallel from a reliability point of view)

c Thermal images of the SSFCL without redundant switch

## 7. Cost Analysis

The MTTF of non-redundant FCL is achieved simply by reversing the failure rate of two switches. In general, this equation can be expressed in the form of:

$$MTTF_{non-red} = \frac{1}{\lambda_{SW}} = \frac{1}{2\lambda_S} \quad (32)$$

According to the equations (8)-(10), compared to non-redundant FCL, a potential benefit brought by the redundant FCL is the possibility of achieving high reliability. So, compared to non-redundant FCL, cost of outage, regarding the customer interruption cost is lower for redundant FCL. Nevertheless, redundant design means the need for more components. So, the investment cost of the redundant FCL is higher than non-redundant FCL. To find the configuration with the lowest cost per million hours, a levelized indicator for each structure with respect to the structure total cost is considered. The levelized cost (LC) of the redundant and non-redundant structures determines the net cost of FCL for expected lifetime divided by its expected lifetime. Consequently, it can be a good measure to compare the economic justification of the redundant and non-redundant structures. The levelized cost of a structure can be expressed as:

$$LC = \frac{C_{tot}}{MTBF} = \frac{C_{inst} + C_{loss} + C_{repair} + C_{outage}}{MTTF + MTTR} \quad (33)$$

Where, $C_{tot}$, $C_{ins}$, $C_{loss}$, $C_{repair}$, $C_{outage}$, MTBF, and MTTR are the net cost of FCL for expected lifetime, investment cost, loss cost, repair cost, outage cost, mean time between failures, and mean time to repair, respectively. The investment cost includes the installation cost that is proportional to the rating of the FCL. The investment cost of an $I^{KA}$ FCL with X redundant switches is:

$$C_{inst} = X \times C_0 \times I \quad (34)$$

Where, $C_0$ is the cost of FCL per ampere. According to the survey of energy information administration (EIA) [29], power loss cost $C_{l0}$ is assumed to be $12^{¢/kWh}$. Therefore, the power loss cost for expected lifetime of FCL is derived as:

$$C_{loss} = P_{loss,x} \times MTTF \times C_{l0} \quad (35)$$

Repair cost consists of the cost of reinstallation, labor and transportation of the technical support staff. The labor transportation cost per failure $C_{LT}$, is estimated to be $300^{\$/kA/day}$ [30]. The repair cost is as follows:

$$C_{repair} = C_{inst} + (C_{LT} \times MTTR) \quad (36)$$

Finally, the outage cost is the total revenue loss from customers because of their inability to access network during the outage period. This cost is significantly higher than electricity prices and related to the system scale. The outage cost per day $C_{d0}$, for a $50^A$ system is around $1200^{\$/day}$ [31]. The total outage cost within expected lifetime of FCL is calculated as follow:

$$C_{outage} = C_{d0} \times MTTR \quad (37)$$

In this paper, the field data, MTTR, and price coefficients, which are listed in Table 2, come from the following sources: the military handbook [26] and the provided field data in [29]-[33]. Here, we consider the previous cases in reliability analysis as an example. The power losses of each thyristor in shunt redundant configuration and without redundant configuration are $0.34^W$, and $1.35^W$, respectively. The results of levelized cost calculation under different cases of redundancy are shown Table 1. As indicated in Table 1, the levelized cost is minimized for shunt redundant configuration (parallel from a reliability point of view). Therefore, shunt redundant configuration (parallel from a reliability point of view) not only improves the reliability level but also reduces the levelized cost of system.

The LC can be easily influenced by dataset. So, it should be noted that the objective of this section was not to judge the economic merits of one configuration over another, but rather to present a method of applying a usage model for comparing different configurations.

**Table 1** Levelized cost of FCL under different configurations

| Configuration | Levelized costs ($/million hours) |
|---|---|
| Non-redundant | 25578.06 |
| Shunt redundant (parallel from a reliability point of view) | 3861.39 |
| Shunt redundant (standby from a reliability point of view) | 6662.11 |
| Series redundant (parallel from a reliability point of view) | 8936.58 |
| Series redundant (standby from a reliability point of view) | 6662.11 |

## 8. Conclusion

A comprehensive reliability analysis between the four possible configurations for the redundant design in component-level (i.e. shunt or series redundant switch with parallel or standby operation from a reliability point of view) is carried out. Proposed analysis provides valuable information to enhance system reliability, to choose better redundant system configuration, and to realize maximum benefit of redundant design.

It is shown that the more reliable configuration is determined by junction temperature of switches. Junction temperature range is formulated in which the SSFCL with shunt redundant configuration (parallel from a reliability point of view) is more reliable. If junction temperature was not in the mentioned range, the SSFCL with series redundant configuration (standby from a reliability point of view) is more reliable. Also, this comparison shows that the SSFCL with series redundant configuration (parallel from a reliability point of view) has the lowest reliability. Sensitivities of more reliable configuration to ambient temperature, power losses, and the thermal resistance of heat sink are analyzed. Thermal resistance of heat sinks is regulated to ensure that shunt redundant configuration with parallel operation from a reliability point of view is more reliable. Also, the optimal configuration was found economically. In addition, to calculate the junction temperature, switches loss calculation are investigated in appendix.

Finally, further research may investigate the redundant design in other levels (i.e. leg-level, module-level and system-level). Since, it is under the research and industry requirement to identify the more reliable configuration.

## 10. Appendices

### 1. Power Loss Analysis

Power dissipation of a semiconductor switch consists of two parts: conduction loss and switching loss. Switching loss is composed by power losses during turn-on and turn-off switching transitions and the conduction losses are the power losses caused by the switch on-state resistance and forward voltage drop. So, the power loss can be derived as:

$$P_{loss} = P_{SW} + P_{Cond} = f_{SW}(E_{ON} + E_{OFF}) + \frac{1}{T}\int_T dE_{Cond} \quad (38)$$

$$E_{Cond} = (V_0 + R_S i(t))i(t).T_S \quad (39)$$

### 2. SSFCL and System Parameters:

**Table 2** SSFCL and system parameters

| Symbol | Quantity | Value |
|---|---|---|
| $L_M$ | Inductance of fault current limiting inductor | $0.02^H$ |
| - | Power switch type | BT151 |
| $R_{th,jC}$ | Thermal resistance between the junction and case | $1.3^{°C/W}$ |
| $R_{th,CA}$ | Thermal resistance between case and ambient (in free air) | $58.7^{°C/W}$ |
| $V_S$ | Secondary side voltage of isolation transformer | $63^V$ |
| ω | Grid angular frequency | $2\pi \times 50^{rad/S}$ |
| $P_L$ | Power of the downstream load | $200^W$ |
| $T_a$ | Ambient temperature | $25^{°C}$ |
| *MTTR* | Mean time to repair [32] | $24^{day}$ |
| $C_0$ | Cost of FCL per capacity [33] | $710^{\$/KA}$ |